# Electronic Structure, Optical and Dielectric Spectroscopy Study of TbMnO$_3$

V. A. Trepakov[1,2*], A. G. Dejneka[1], O. E. Kvyatkovskii[2], D. Chvostova[1],
Z. Potucek[1], M. E. Savinov[1], L. Jastrabik[1], X. Wang[3] and S.-W. Cheong[3]

[1] Institute of Physics, AS CR, Na Slovance 2, 182 21 Prague 8, Czech Republic
[2] Ioffe Institute of the RAS, 194 021 St.-Petersburg, Russia
[3] Rutgers Centers for Emergent Materials and Department of Physics and Astronomy
  Rutgers University, NJ 08854

[*] e-mail: trevl@fzu.cz;

**ABSTRACT.**

Spectral ellipsometry studies of optical functions in the region of 0.75-8 eV, first observation of photoluminescence, and low-frequency (100 Hz–1 MHz) dielectric permittivity investigations were performed in ferroelectric-multiferroic TbMnO$_3$ single crystals. The main attention was paid to electronic band structure, elucidation of the nature of the main optical transitions and unusual behavior of the dielectric permittivity in the region of the sharp narrow $\varepsilon_c'(T)$ maximum associated with ferroelectric phase transition ($T_C$ = 27.4 K). It was found that magnitude of the $\varepsilon_c'(T)$ maximum decreases strongly with frequency, whereas position of maximum is frequency independent. It was shown that observed features of polarization response can be described satisfactorily on the basis of the Landau-Khalatnikov theory for critical relaxation polarization.

**INTRODUCTION.**

Remarkable family of orthorhombically distorted perovskite-like rare earth manganites $R$MnO$_3$ (R = La, Pr, Nd, Sm, Eu, Gd, Tb, Dy) has been attracting a plenty of attention exhibiting new diverse properties, of high fundamental academic and application interests (such as colossal magnetoresistance [1, 2], etc). Among of them, manganites with R = Eu, Gd, Tb, Dy are turned out to be the objects of the peculiar interest. At room temperature they are paramagnetic ones. Below 30-40 K antiferromagnetic ordering appears followed by induced long order ferroelectric (FE) phase transitions (PT) [3-6]. Close proximity of FE and magnetic transitions provides strong coupling between magnetic and FE degree of freedom, intricate interplay of the lattice, charge, orbital, and spin, which lead to emergence of unique properties and effective magnetic control of spontaneous polarization [3]. In this regard, orthorhombic TbMnO$_3$ (TMO) has attracted a special considerable attention. Namely studies this material led to discovery of spin-spiral-driven-ferroelectric-multiferroics (*SSDMF*) revealed efficient switching phenomena [3, 6-11]. At room temperature and ambient pressure TMO exhibits the central inversion orthorhombically distorted perovskite-like structure (*Pbnm* ($D_{2h}^{16}$, N62 space group, *mmm* point class, Z = 4, 20 atoms per unit cell; the lattice parameters $a$ = 5.293 Å, $b$ = 5.838 Å, $c$ = 7.403 Å), containing a network of corner-sharing MnO$_6$ octahedral clusters [12] as basic elements of crystalline and electronic structure. Above 41 K, in paramagnetic phase, it is conventionally considered as Mott insulator containing $d^4$ ions (*Mn$^{3+}$ in oxygen octahedral cage*) with orbital degeneracy. Unlike conventional band insulators, in Mott insulators spin and orbital degree of freedom still conserve. Consensual description of magnetic properties of manganites is usually based on the localized spins model in which $t_{2g}$ and $e_g$ states are localized and form total spin $S_i$ = 2 of Mn centers. In such the case, magnetic order is caused by anisotropic Kramers-Andersen exchange over oxygen orbitals. Valence electronic states of MnO$_6$ clusters consist of valence states of respective atoms: fifth $d$ orbitals of Mn ($d_{z^2}$, $d_{x^2-y^2}$, $d_{yz}$, $d_{xy}$, $d_{xz}$) and three O2$p$ orbitals ($p_{x,y,z}$) for each oxygen atoms of octahedrons. In orthorhombic manganites Mn$^{3+}$ ions of $t_{2g}^3 e_g^1$ configuration are located in a distorted octahedral environment of six O$^{2-}$ ions and the Jahn–Teller (J-T) instability occurs in contrast to hexagonal ones. In $e_g$ state a checkerboard order of $d_{3x^2-r^2}$ and $d_{3x^2-r^2}$ orbitals in TMO leads to appearance of several competitive exchange interactions. At $T_N$ ~ 46 K Mn$^{3+}$ ions form a collinear sinusoidal spin wave inducing PT into pure collinear antiferromagnetic structure with an incommensurate sinusoidal spin ordering with $q_{Mn}$ ~ 0.295 with spins oriented along the [010] direction [13]. Herewith, the collinear spin density waves (SDW) does not induce a uniform electric polarization ($P$ = 0). However, at further cooldown the wavelength of SDW $q_{Mn}$ decreases. Below $T_H$ SDW is locked at incommensurate wave vector $q_{Mn}$ ~ 0.28 [8, 13, 14] at which *bc* cycloidal CWD (Mn spins rotate in the *bc* plane) occurs. It breaks the lattice inversion symmetry inducing ferroelectric (FE) phase transition (PT) of the second order ($T_C$ = $T_{loc}$ ~ 28 K) with $P_S$ orientation along *c* – axis, while the antiferromagnetic (AFM) ordering along *b* –axis persists [3]. So, polarization of a pure relativistic nature emerges due to spin-orbit interaction (SOI) which control P-M coupling [7] and the magnitude or the direction of $\bar{P}$ can be drastically changed by applying a magnetic field [3,15]. Below 7 K a quasi-long range ordering of magnetic moments of the Tb$^{3+}$ ions takes place [12, 13]. As result, ferroelectricity of pure electronic contribu-tion due to SOI can appear already in clamped lattice of TMO [16]. Experimental spontaneous polarization $P_S \sim 600$ μC/m$^2$ [3] is small ($\sim 2.6 \times 10^{-2}$ C/m$^2$ at room temperature, that is $10^3 - 10^4$ less than that in BaTiO$_3$), but magnetoelectric and magnetocapacitance effects, which are attributed to switching of the electric polarization induced by magnetic fields, are gigantic [3]. Lowering of the electronic density distribution symmetry leads to lowering of the lattice symmetry and appearance of ionic contribution $P^{ion}$. Herewith, only full polarization $P = P^{el} + P^{ion}$ has correct direction and magnitude corresponding to experiment. Polar distortions of ions, which lead to correct $P^{ion}$ are anomalous small (< 1 mÅ) [17-21] and experimental structure below 27 K is detected as central inversion one [3]. Recent highly precise x-ray diffraction experiment [21] demonstrated that average ionic displacement of the Tb$^{3+}$ ions along c- axis to be -21 ± 3 fm, yielding $P^{ion} = 176$ mC m$^{-2}$, that is a quarter of experimental $P \approx 600$ mVm$^{-2}$.

To the moment two main microscopic mechanisms of ferroelectricity in spiral magnets, *lattice* and *electronic* ones, have been suggested [16-20, 22-26]. In *lattice mechanisms* (e.g. [22-24]) non-collinearity of magnetic moments (inhomogeneous magnetization) induces polar optical displacements conventionally discussed in terms of inverse Dzyaloshinskii-Moriya interaction, which is proportional to SOI. The homogeneous polar order appears when spiral spin structure lowers the symmetry of the system leading to effective magnetoelectric coupling term and emergence of electrical polarization $P \sim e_{ij} \times (S_i \times S_j)$, where $S_i$ and $S_j$ are the neighboring spins and $e_{ij}$ is the lattice vector connecting them. Pure *electronic mechanisms* consider emergence of electronic polarization due to SOI [16-20, 26, 27] and, e. g., estimation of spin-orbit contribution performed in [19] yields $P \sim 32$ μC/m$^2$. Consideration of spin-orbit and ionic contributions leads to value of $P \sim 467$ μC/m$^2$ which agrees well with experimental value [3].

Nature of electronic states, role of electronic contribution into ferroelectric polarization in *SSDMF* materials is turned out to be one of the main problems in physics of manganites and ferroelectromagnets of the day. But its solution is strongly hampered by lack of reliable experimental and theoretical data on electronic states, band structure, electron-phonon and magneto-electric interactions interaction, and their temperature evolutions. Even for TMO, model canonical prototype ("guinea pig") for SSDMF, the few published experimental and theoretical studies (e.g. [20, 28-33]) have been done in the rather rough approximations, have not been supported by essential experiments and need further development.

In this paper, we present detail spectral ellipsometric studies of the optical constants spectra (0.75-8 eV) for determination of energy gap, main optical transitions, features of the electronic band structure, as well as the first observation of photoluminescence in TMO crystals. Obtained results and their interpretations were made taking into account the most substantive results of experimental studies and theoretical simulations available. Also we report on and shortly discuss surprising minimum of the surface roughness was found at T ≈ 126 K. A special attention was paid to low-frequency (100 Hz–1 MHz) dielectric permittivity study in the region of unusually sharp and narrow $\varepsilon'_c(T)$ maximum associated with FE PT ($T_C$ = 27.4 K). It was shown that unusual narrowness of $\varepsilon'_c(T)$ maximum and specific dispersion observed are a sequence of pronounced magneto-electric coupling which are discussed in the frame of the Landau-Khalatnikov theory for critical relaxation polarization [34].

**EXPERIMENTAL**

Single crystals of TMO have been grown by the floating zone method in Rutgers University. Structure identification and perfectness (single crystallinity) were controlled by x-ray diffraction and exhibited orthorhombic *Pbmn* single crystalline structure without impurity phase. Experimental specimens were oriented using Laue XRD patterns and prepared as polished plane-parallel plates of dimension 3 x 2 x 2 mm$^3$ oriented along ***c***-, ***a***- and ***b*** axes associated with <100> directions of the pseudocubic $O^1_h$ phase with the largest ***bc*** plane. Magnitude of resistivity at Room temperature tested by four-probe method was amounted as ~ 175 kΩcm that is consistent for Mott insulators. Thermo-electromotive force measurements revealed *p*-type of conductivity. Ellipsometric spectra of optical constants ***n*** and ***k*** were measured using a variable angle spectroscopic ellipsometer J.A. Woollam operating in a range of photon energies $E$ = 0.75–8 eV (λ ~ 160–1600 nm) within the temperature region 4.2 – 300 K. The main ellipsometric angles ψ and Δ were measured in the reflection mode at angle of incidence 65°–75°. To eliminate depolarizing effects of back surface reflection from the substrate the back surface of the substrates was mechanically roughened. The ellipsometric spectra were analyzed with the WVASE32 software package considering surface roughness. Surface roughness of experimental specimens was ~2 nm or less. Perturbation of the ellipsometric data was mathematically removed by assuming that the surface roughness could be approximated as a Bruggeman effective medium [35]. For spectral dependences of optical constants the generalized multi-oscillator model was used. Usage of an optical cryostat in temperature measurements brought an insignificant error in measurement of optical constants strictly along basic axes. The photoluminescence emission spectra of the crystals fixed to a copper holder of a closed-cycle helium refrigerator were recorded at 12–300 K using a set-up based on a McPherson 2061 1-meter focal length grating monochromator. Photoluminescence intensity in the 1.55 - 4.13 eV (300–850 nm) of spectral region was measured with a cooled RCA 31034 GaAs photomultiplier operating in the photon-counting mode. Photoluminescence was excited with light from a high-pressure Xe lamp filtered through a double-grating Jobin-Yvon DH 10UV monochromator or with a line of HeCd laser (442 nm, 2.8 eV). All emission spectra were corrected for the spectral dependence of the apparatus response. The complex dielectric permittivity $\varepsilon_c = \varepsilon' + i\varepsilon''$ was measured under an *ac* reference field of ~30 Vcm$^{-1}$ in the 100 Hz–1 MHz frequency region with a 4192 LF Hewlett-Packard Impedance Analyzer assembled with a He-flow cryostat allowing temperature cycling within 7–300 K with a rate 10–20 mK s$^{-1}$. Detailed measurements in the region of the narrow $\varepsilon_c'(T)$ maximum were carried out with the temperature satiation ± 15 mK. Pt–Au electrodes were evaporated onto opposite *ab* plane faces of thin TMO condenser platelets, whereby reference electric field was applied along the *c* axis.

**RESULTS AND DISCUSSIONS.**

**Optical Constants, Electronic Structure, Photoluminescence.**

Figures 1, 2 present room temperature spectra of the real and imaginary part of the optical dielectric constant ($\varepsilon' = n^2 + k^2$; $\varepsilon'' = 2nk$) extracted from ellipsometry in *a*-, *b*- and *c*-axes polarization. The pronounced dichroism of the dielectric functions characterizes the optical anisotropy of TMO. Presence of *A*-, *B*-, *C*- and *D* optical absorption bands in the region of ~ 1.7 – 2.7 eV, 3.7- 3.8 eV, 5-5.2, and in the region of ~ 7.00 eV respectively is displayed in the Fig. 2. Surprisingly, it was turned out to be that visible optical losses extend rather far into IR region revealing unknown structure in the region of 0.7 – 0.87 eV. The absorption coefficient $\alpha = 2\omega k/c$ determined at room temperature from the ellipsometry measurements is shown in Fig. 3. Figure 4 shows a Tauc plots for direct optical transitions $\alpha^2 \sim h\nu - E_d$ for light polarization along *a*- and *b*- axes. Energies of the primary direct optical transitions ($E_d$) determined from the intercept of a linear fit of $\alpha^2$ plot versus energy $h\nu$ were found to be 3.1 and 3.2 eV along *a*- and 3.0 and 3.72 eV along *b*- *a*xis. At the same time, for both polarizations the absorption coefficient below 3 eV can be fit by the direct optical transitions expression too and gives ~ 1.7 eV for the energy of the lowest direct transition.

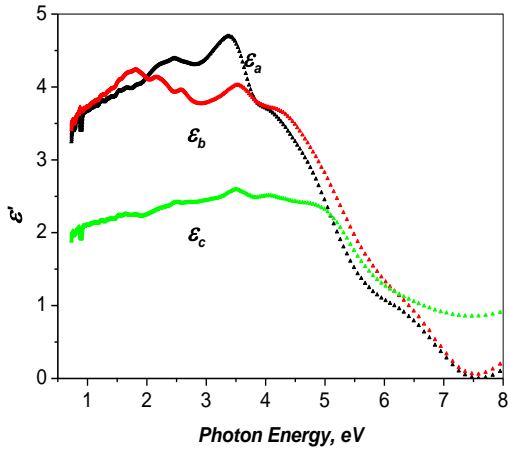

Fig. 1. Real part of optical dielectric constant in *a*-, *b*- and *c*- axes polarization of TbMnO$_3$ single crystals (room temperature).

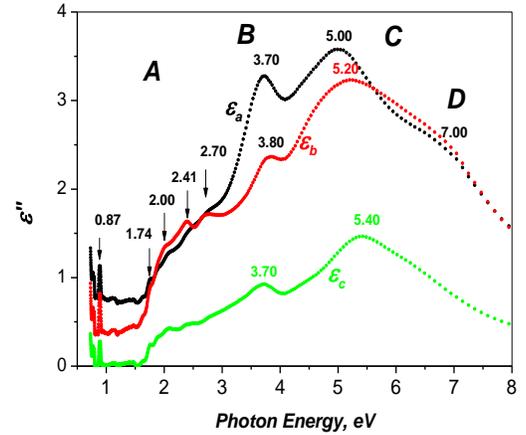

Fig. 2. Imaginary part of optical dielectric constant in *a*-, *b*- and *c*- axes polarization of TbMnO$_3$ single crystals (room temperature).

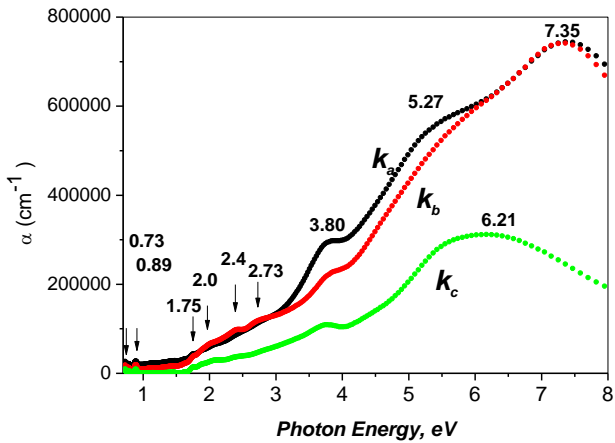

Fig. 3. Absorption coefficients spectra for *a*-, *b*- and *c*- axes polarization of TbMnO$_3$ single crystals.

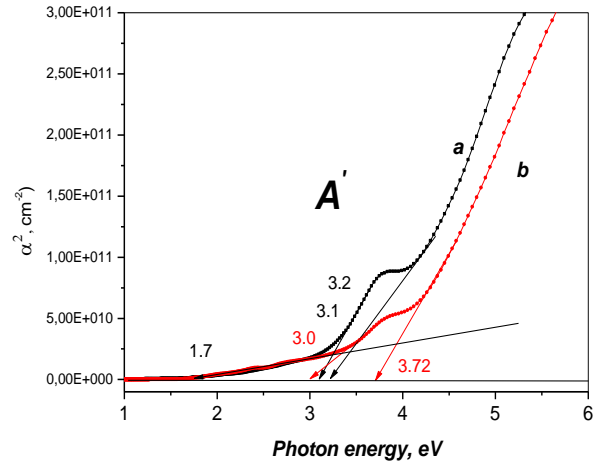

Fig. 4. Tauc plot for direct optical transitions energies for light polarization along *a* and *b* axes.

Figures 5 and 6 present the Tauc plot $\alpha^{1/2} \sim h\nu - E_d$ for indirect and direct optical transitions for the light polarization along *c*- axis. It is seen that the lowest in energy *indirect optical* transitions situated at $E_{i1}= 0.7$ eV (up to phonon energy accuracy) and dominate until ~ 3.7 eV. At higher energies the more intensive indirect transitions with $E_{i1} = 2.17$ eV prevail. In the region above ~ 4 eV the stronger direct optical transition with the energy of ~ $E_d = 4.5$ eV begin to dominate.

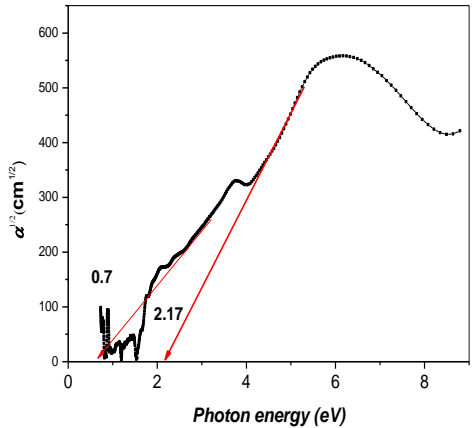
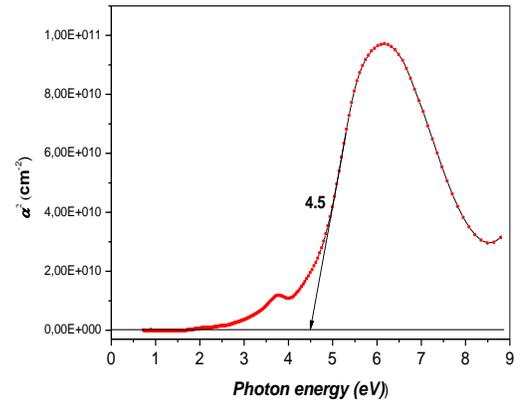

Fig. 5. Tauc plot for indirect optical transitions for *c* – polarized light.

Fig. 6. Tauc plot for direct optical transitions for *c* – polarized light.

Figure 7a,b shows details of the absorption spectrum in the IR region. It consists of two slightly resolved doublets at 0.77 and 0.79 eV and another, stronger one, at 0.88 and 0.89 eV. The small absorption maximum at ~ 0.85 eV presents too.

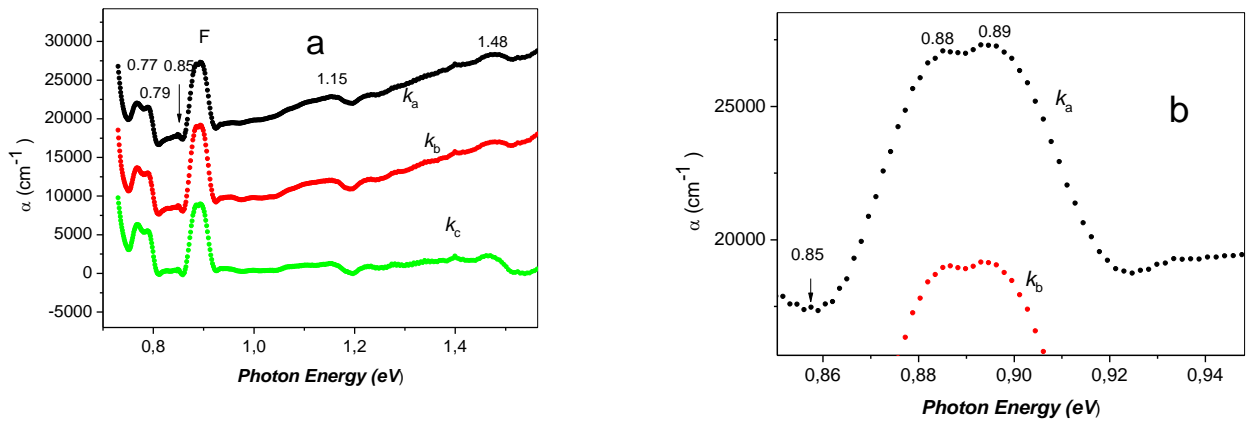

Fig. 7: a) IR optical absorption in IR region for *c*- polarized light; b) details of IR absorption spectra

Table 1 summarizes characteristic energies of the optical absorption spectra of TMO single crystals. Figure 8 presents spectra of the absorption coefficient were taken at different temperatures for light polarized along *c* axis.

**Table 1. Characteristic energies in optical constant, absorption and luminescence spectra (eV)**
*Subscripts label polarization of light and type of interband optical transitions*

| Bands | ε″ | α maxima Figs. 3,7,8 | Direct transitions along a,b Fig. 4 | Transitions along c Figs. 5,6 | Luminescence maxima | |
|---|---|---|---|---|---|---|
| | | | | | Emission | Excitation |
| | | | | 0.7 *ind* | | |
| G | | 0.77 – 0.79 | | | | |
| | | 0.85 | | | | |
| F | 0.87 | 0.88 – 0.89 | | | | |
| | | 1.15 | | | | |
| | | 1.48 | | | | |
| A | 1.74 | 1.75 | 1.7 *a,b* | | | |
| | 2.00 | 2.0 | | 2.17 *ind* | 2.28; 2.51; 2.67 | |
| | 2.41 | 2.4 | | | | 2.60; 2.66 |
| | 2.70 *a,b* | 2.73 *a,b* | | | | |
| A′ | | | 3.0 *b* | | 3.0; 3.19;– 3.37 | |
| | | | 3.1 *a* | | | |
| | | | 3.2 *a* | | 3.37 | 3.39 |
| B | 3.70 | 3.80 | 3.72 *b* | | | 3.85 |
| | | | | 4.5 *dir* | | 4.30 |
| C | 5.00 *a* | | | | | |
| | 5.20 *b* | 5.27 *a* | | | | |
| | 5.40 *c* | | | | | |
| | | 6.21 *c* | | | | |
| D | 7.00 *a,b* | | | | | |
| | | 7.35 *a,b* | | | | |

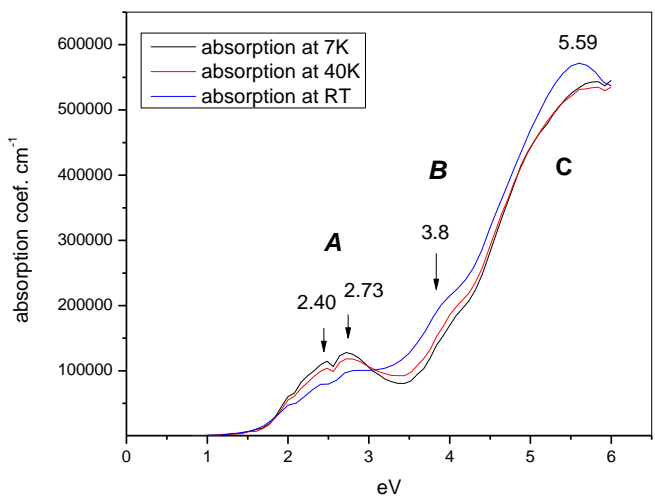

Fig. 8. Optical absorption spectra in ***c*** axis polarization.

It is seen that **B** and **C** absorption bands shift distinctly to higher energies at cooling in consistence with conventional behavior of the band edge transitions. It is generally assumed that structured **A** absorption band is connected to intersite $e_g^1 \to e_g^2$ transition across the Mott gap [37, 38]. According to [36-38] this so-called "2 eV absorption" is pronounced well in manganites with relatively large ionic radius of *R* ions and large Mn-O-Mn bond angle $\phi$ (NdMnO$_3$, PrMnO$_3$ and LaMnO$_3$) and absent in GdMnO$_3$ and TMO with smaller *R*-ions and bond angle [36-38]. Figure 8 shows, however, that **A** absorption band is quite pronounced in TMO at room temperature and markedly increases at low temperatures (at least at 40 K and 7 K) that can be connected to conventional temperature effect, low temperature AFM and FE PT and $\phi$ angle increasing.

Figure 9 presents unpolarized luminescence spectra of TMO single crystals, which distinctly emerges below 200 K under excitation at $\lambda_{ex}$ = 310 nm (4 eV). It is seen that the main wide emission bands locate in the region of **A**, **A′** optical transitions between 1.5 – 3.5 eV. At "high temperatures" above transition point between incommensurate antiferromagnetic and paramagnetic phase ($T_N \approx 46$ K), luminescence spectrum consists of the dominated structured **A′** emission band in the ~ 2.7 – 3.5 eV range and the shoulder at ~ 2.55 eV in the **A** optical transitions region.

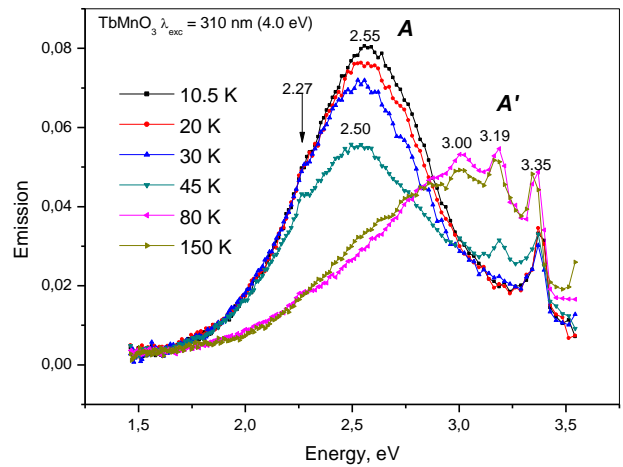

Fig. 9. Emission spectra under excitation at wavelength 310 nm (4 eV).

In is seen that emission spectrum changes strongly at cooling. Intensity of **A′** band decreases as well as its structure pronuncation. At T ≤ 30 K only one relatively narrow emission line of **A′** band centered at 3.35 eV remains. In contrast, intensity of *A* band, centered at ~ 2.50 – 2.55 eV, increases and a structure emerges at ~ 2.37 eV. Figure 10 shows that at 10.5 K emission spectrum can to be successive described by four Gaussian centered at 2.28, 2.51, 2.67 eV (*A* band) and narrow line at 3.37 eV (*A′* band).

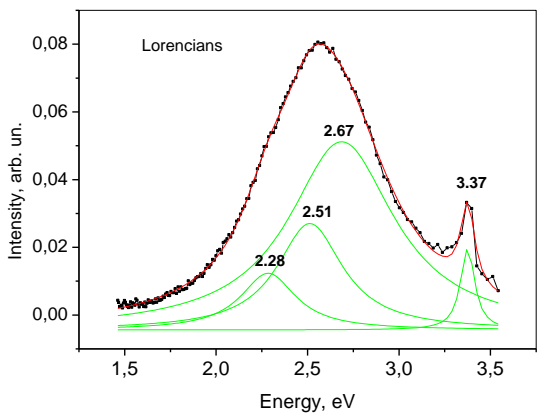
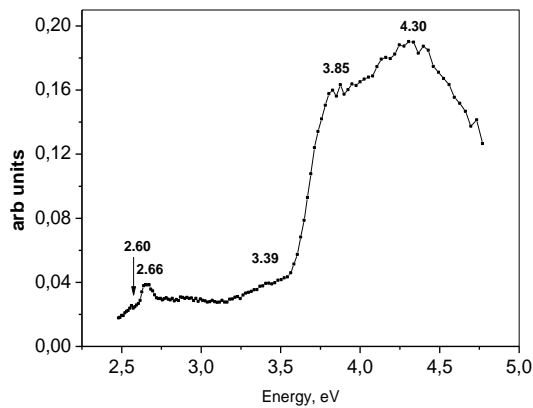

Fig. 10 Four Gaussians fit; T= 10.5 K

Fig. 11, Excitation spectrum for emission registered at 550 nm (2.25 eV); T = 11 K

Figure 11 presents luminescence excitation spectrum for emission registered at $\lambda_{em}$ = 550 nm (2.25 eV, i. e. at the maximum of the *A-* band) was taken at 11 K. It is seen that excitation spectrum consists of the rather weak maxima in the region of *A-* and *A′* bands (2. 60 and 2.66 eV), the shoulder at ~ 3.39 eV and two strong maxima at ~ 3.85 eV and 4.30 eV in the *B*-band region. The energies of the luminescence spectra maxima are given in the Table 1.

Starting discussion of the obtained results let us define the basis and the main theoretical calculations and experimental results available in comparison with which it can be done. Unfortunately, despite many year studies, a lot of principal questions concerning electron band structure and nature of the optical transitions in manganites remain open. Reported theoretical and experimental results and conclusions are rather inconsistent and contradictory still. Today, apparently, the most studied and known is electronic structure of orthorhombic $LaMnO_3$ (LMO), the parent compound of the colossal magnetoresistance manganites (see. e.g. [39-46] and references therein). But due to presence of *f*- electrons, which are absent in LMO, electronic structure of TMO is a significantly more complex task and calculation of its electronic structure has not been perfor-med properly enough up to now. Together with Jahn-Teller distortions, playing important role in electronic structure, there are various complications with this system that one must be aware of and which can explain the absence of reliable theoretical calculations in the literature. This very complicates interpretation of the experimental optical spectra. For example, it is hard to model realistic cycloidal spin structure, although for band structure one could possibly to get it away with simple antiferromagnetic ordering. It is necessary to treat correctly the *d* and *f* electrons and, properly speaking, a reliable way to take into account the Tb 4*f* electrons is not clear. These are very strongly correlated electrons and DFT fails to describe them properly. The spin-orbit coupling must be taken into account too. Since effects of SOI are usually small, the calculations must be done with high-precision requirements for both the total energy in the electronic structure and the forces for the structural relaxation of the system. To model the spin spiral itself, one has to use large supercells etc. As up to day there are only three relatively successive theoretical publications on electronic structure of orthorhombic TMO [20, 30, 32], each of they had been executed at some simplifications and approximations. So, Malashevich [20] used the plane-wave pseudopotential method based on DFT method with on-site Coulomb interaction LDA-U and GGA-U approximations implemented within the VASP code [47] with PAW potentials [48,49]. But the Tb potential was considered without contributions of *f* electrons in the valence. It was found that LDA predict TMO to be metallic in the ground state. However, usage of the LDA+U method shows that already at small values of *U* a finite band gap appears, and the system becomes insulating. The magnitude of the band gap of the order of 0.5 – 1 eV has been obtained at the value of Coulomb interaction between *d* electrons in the Hubbard-like fashion (on site Coulomb repulsion) U = 1 eV without usage of the scissor operator. Since the periodic boundary conditions are required by superlattice approach, results [20] could not model spin spirals with arbitrary wave vector, but only spirals that are commensurate with the crystal lattice. As result, Malashevich [20] successive calculated electronic structure for cycloidal spin structure of TMO but consider *f* – electrons of Tb atoms at all (and as a sequence neglecting SOI) for commensurate crystal lattice only.

More recently first principle calculations of optimized rhombohedral distorted TMO have been performed [30,32] within CASTEP code [50]. The Vanderbilt ultrasoft pseudopotential with valence electrons correspond to $Tb4f^85s^25p^25d^16s^2$, $Mn3d^54s^2$ and $O2s^22p^4$ was used to describe the electron-ion interaction. As in [19] the GGA approximation in the Perdew-Burke-Ennzerhof scheme was employed to describe the exchange correlation functional [51, 52]. Factually, the calculations [30, 32] are standard GGA calculations, i.e. the bands obtained are quite similar to those obtained by usual LDA. Usage of scissor operator in [30, 32] suggests that band gap has been adjusted, but it is unclear with what the accuracy it had been done. So, theoretical results [20, 30, 32] provide a possibility to use them proposing tentative interpretation our optical spectra in orthorhombic state of TMO, where theoretical simplifications used are not too dramatic allowing us to offer acceptable qualitative picture. Figure 12 shows electronic band structure around the Fermi energy $E_F$ received in [32] and recomposed using the scissor operator of 3.0 eV, which fits experimental value of band gap ~ 0.5 eV received from conductivity temperature dependence of TMO epitaxial thin films [28].

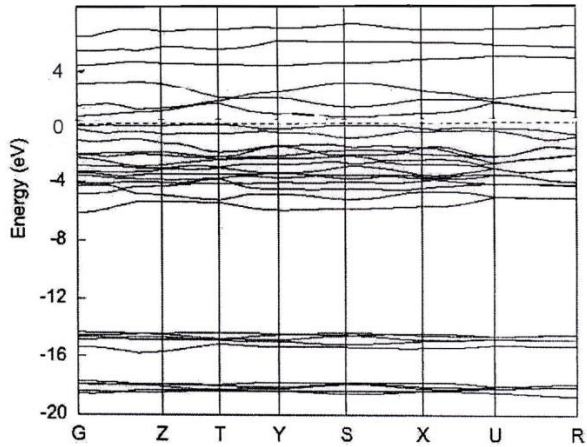

Fig. 12 electronic structure [32] recomposed by usage of the scissor operator of 3.0 eV

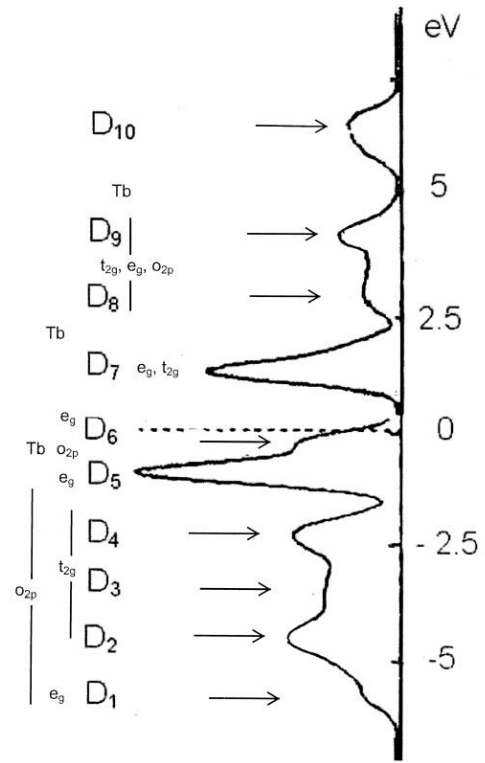

Fig. 13. The tentative DOS structure on orthorhombic TMO was composed on the base of [42, 20] calculations revision and our experiments.

Figure 13 shows tentative DOS structure on orthorhombic TMO was composed on the base of [41, 42] calculations revision and our experiments. Following [20, 32] the electronic structure of the top of the $D_1 - D_6$ *valence bands* (*VB*) is controlled mostly by O 2p Mn 3d ($t_{2g}$), and Tb f states. In the region of ~ 0 - 1.47 eV ($D_5 - 0$ eV) DOS is rather small and electronic states are composed of nearly equal contributions of Mn $e_g$ dominating in $D_5$ band and O 2p ($D_6$) levels [20]. In the range of

– 6.35 ÷ – 1.47 eV DOS is composed mostly of the O 2$p$ (D1-D4) [32] with an admixture of Mn $t_{2g}$ levels [20]. The bottom of the partially occupied conduction bands (*CB*) are originated mainly from Mn 3$d$ $e_g$ and Tb 4$f$ [32] with small admixture of O 2$p$ levels [20]. According to [20] it consists of two sub-bands. The first one has a rather small DOS maximum at ~ 1 eV consisting of dominating Mn $e_g$ levels together with O 2$p$ admixtures. In the region of ~ 2.2 – 3.3 eV the strong maximum of DOS takes place, which is constructed from dominating Tb $f$ levels [32] with possible addition of Mn $t_{2g}$ states [20]. In the 5 – 7 eV range the region of pronounced DOS of *CB* presents. It is originated mainly from Mn $t_{2g}$ states. Regarding nature of the band gap transitions, according to [32] the top of *VB* locates at the **T** point of Brillouin zone (BZ) and the minimum of the lowest CB locates at **S** and **Y** points. In such the case band edge spectrum is originated from **T** → **S** and **T** → **Y** optical transitions. At the same time, according to [20], the top of VB as well as the bottom of CB was found at the $\Gamma$ point of BZ and band gap is originated from direct $\Gamma \to \Gamma$ optical transitions.

Following this scheme of the energy structure, and taking into account results [20, 32] we suggest that the ***G*** and ***F*** narrow absorption doublets recorded by us at 0.77, 0.79 eV and 0.88- 0.89 eV respectively (Table 1) are originated from band edge interband transitions between maximums of DOS of *VB* and *CB* closest to Fermi energy. According to scheme [32] and Fig. 13, they can belong to transition between the nearest to $E_F$ shoulder of (DOS) at **D₆** region to the bottom of the *CB*, i e, due $p$-$d$ charge transfer from O 2$p$ bands to Mn $e_g$ subband [20]. It should be pointed out that ***G*** and ***F*** absorption are unlikely insulating gap transitions. Absorptions between ~ 1.5 and 3.3 eV (***A*** and ***A*′** absorption bands and $\varepsilon''$ maxima) are caused be optical transitions between **D₅** ($d$ $e_g$) to **D₇** ($f$, $t_{2g}$ and $e_g$) DOS bands, i, e, by charge transfer from Mn $d$- to $f$ states of Tb and $d$-$d$ electronic transitions within Mn$^{3+}$ ions. Thereupon, optical transition at 2.7 eV has to be considered as the indirect fundamental one.

Presence of ***B*** band (Fig. 2) can be associated with transitions between **D₄** and **D₇** DOS bands originating from charge transfer from O 2$p$ to Tb and Mn $f$ and $d$ states. At last **C** and **D** bands can be associated with $p$-$d$ charge transfer electronic transitions (**D₁, D₂, D₃, D₄**) → (**D₈, D₉**), mostly between O 2$p$ and Mn $d$ states. Transition at 4.5 eV should be considered as the major direct fundamental ones at the band edge .

Luminescence excitation spectrum of Fig. 11 consists of two moderately pronounced bands peaked at 2.56 – 2. 66 eV and at 3.39 eV which can be treated as electronic charge transfer between **D₆** (O 2$p$) to **D₇** ($f$, $t_{2g}$ and $e_g$) and transition between **D₅** ($d$ $e_g$) to **D₇** ($f$, $t_{2g}$ and $e_g$) DOS of Mn and Tb ions. The most intensive excitation bands with maximums at 3.85 eV and 4.20 eV can be connected with charge transfer from O $2p$ to $Mn^{3+}$ states. In spite of luminescence emission reveals in the region of *A* and *A′* optical transitions (Table 1, **D$_7$** → **D$_5$**) its detail interpretation is a rather complex task and can include several possibilities. Conventionally, emission is originated by transitions of relaxed electrons from the bottom of *CB*. So, two complex emission bands can be produced by recombination of photoexcited electrons from the lowest *CB* with holes of O $2p$ of **D$_2$**, **D$_3$**, **D$_4$** bands. Also, contribution of $^5T_2 \to {}^5E$ transitions of $Mn^{3+}$ ions as well as appearance of $Mn^{4+}$ centers under illumination providing emission originated from $^4T_2 \to {}^4A_2$ and $^4T_1 \to {}^4A_2$ transitions [53] are possible too. Surprisingly, emission maximums at 2.38 eV and 2.51 eV (Fig. 10) coincide with position of characteristic emission lines of *f-f* transitions $^5D_4 \to {}^7F_6$ and $^5D_4 \to {}^7F_5$ of $Tb^{3+}$ centers (e.g. [54]). Such centers present in TMO initially and under photoexcitation by light their broadened emissions lines can manifest themselves in contour of complex overlapping contributions into luminescence spectra. Strong changes of the luminescence spectra with temperature can be connected to Mn centers and TMO structure reconstructions accompanying PTs.

Obtaining of detailed data on optical absorption and luminescence spectra of TMO single crystals, their temperature evolutions and interpretation of the main optical transitions was performed on the basis of suggested updated electronic band and DOS scheme can be considered as one of the important results of presented work. Unknown earlier structured absorption was found in the near IR region and its interpretation was suggested. The minimal optical gap of TMO was found to be ~ 0.75 eV for indirect optical transitions.

**Thermooptics and Surface Roughness**

As it was pointed out, determination of electronic contribution into FE PT and its role in formation of the electric polarization is one of the most principal and interesting problems of TMO investigations. At the first glance, one can to believe that a noticeable anomaly or pronounced feature of the temperature dependence of the low-frequency electronic polarizability can present in the region of FE transition. Due to limitation of the spectral region accessible in our experiment, in order to try to define character of the temperature behavior of the "static electronic polarizability" in the FE transition region we undertook detailed ellipsometric study of the temperature evolutions (0.5 K steps) of the real part of the optical dielectric function in the region of 23 – 42 K for light polarized along *c* axis. Figure 14 presents respective spectra obtained.

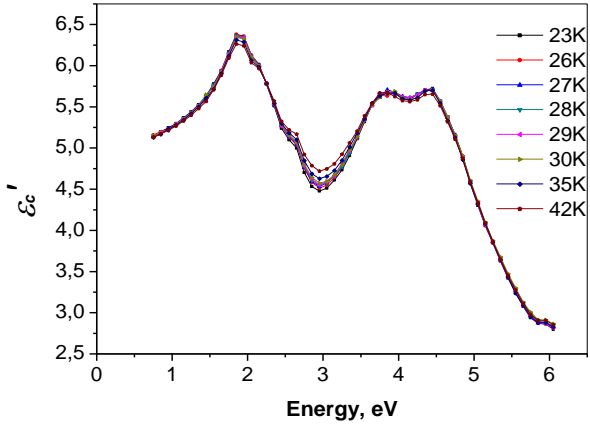

Fig. 14. Real part of the optical constant is taken at different temperatures in the region of FE phase transition.

It was found that actual refractive index spectra are described well by Sellmeier formula with four

$$n^2 - 1 = S/[1 - (\lambda_s/\lambda)^2], \qquad (1)$$

effective Lorentzians. Using the wasf_3.0 code and describing spectral dependence of the optical constants by the generalized multioscillator model for four oscillators we estimated magnitude of

$$\varepsilon^*(\omega) = \varepsilon_\infty + \sum_{j=1}^{n} \frac{\Delta\varepsilon_j \omega_j^2}{\omega_j^2 - \omega^2 + i\omega\gamma_j} \qquad (2)$$

optical dielectric permittivity at zero frequency $\varepsilon'_{el}(0)$. However, such procedure did not elicit any visible temperature dependence of the $\varepsilon'_{el}(0,T)$ permittivity in the region of 23 − 42 K, where its magnitude was amounted as $\varepsilon'_{el}(0) \approx 5.7$. It means that either possible feature of the temperature behavior of the electronic polarizability in the region of FE PT is smaller than used procedure error, or it is absent. Another attempt was made studying temperature dependences of the refraction indices (thermooptics) at selected energies of the light photons. It have been found that for energies in the region of $A'$- $B$, $C$ and $D$- bands $n(T)$ dependence reveals the most general case of positive thermooptic effect ($dn/dT > 0$) with no visible features in the region of FE PT. Whereas temperature dependence of refraction index, for the light with $h\nu = \lambda = 2.5$ eV belonging to *A*- band reveals a pronounced breakpoint and a rather faint local maximum at ~ 27 K (Fig. 15). Because, such effect was observed only at the specified energy of quantum of light, it is difficult to conclude definitely whether the observed effect is connected to small anomaly of electronic contribution into FE transition, or it is a secondary effect of PT. On the other hand, there are no wonder because namely optical absorption in this spectral region is controlled by $d_i - d_j$ *intersite* optical transitions, which are sensitive to both orbital and magnetic ordering.

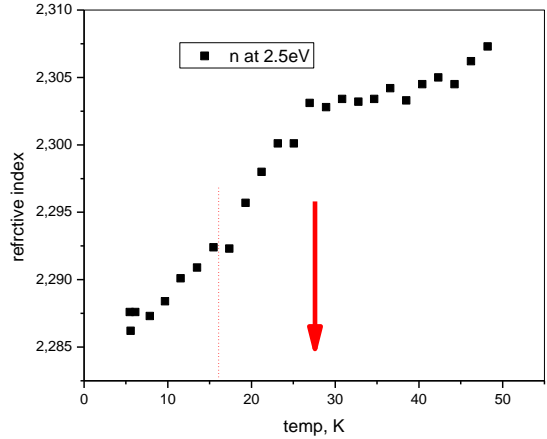

Fig. 15. Refraction index is taken at the energy 2.5 eV (λ = 496 nm)

Absence of noticeable anomaly of the optical dielectric function (i. e. electronic polarizability) in the region of FE transition is a natural sequence of electron-lattice interaction. Besides, simple approximation of the optic permittivity in the actual case is rather doubtful because does not take into consideration electronic dispersion which appears at low frequencies due to interaction with reconstructing magnetic system. In TMO transition into FE phase is caused by namely magnetic spin system whose frequencies much lower than electronic ones and all this is playing at low frequencies very far from measured optical ones. In electronic system at any frequencies, *including static contribution*, only small changes in narrow temperature region of FE phase transition can be realized. So, in optical experiments it is very heavy to study and obtain data about quasi-static (very low frequency) electronic dispersion. But electronic polarizability itself in paraelectric and FE phases is different and some breaks and/or weak changes in *n*(T) dependence can appear, like it was observed in our experiment.

.

Figure 16 presents the temperature dependence of the *ab* surface roughness. A surprising temperature minimum of the roughness was found in the region of ~ 126 K while no any phase transitions are known for TMO above ~ 48 K.

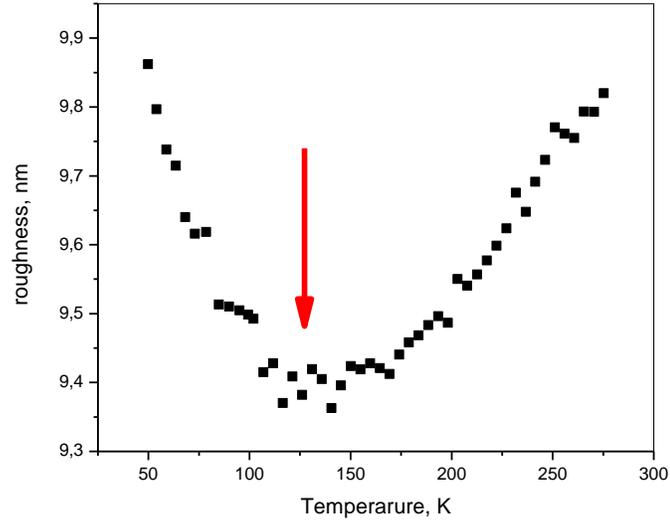

Fig. 16. Temperature dependence of the *ab* surface roughness was taken at light with hν = 2.0 eV.

The nature of such effect demands a scrutiny, and probably can be a sequence of some reconstructtion of TMO surface structure in the region of 126 K. The similar suggestion has been reported in [28], where surprising weak maxima of magnetic susceptibility in TMO thin films on $LaMnO_3$ and $SrTiO_3$ substrates were found at 110 K and 120K respectively. Such features and difference in the temperatures of magnetization maxima was tentatively connected to possible AFM PT due to change of $Mn^{3+}$ magnetic moment and different lattice misfit for TMO films on $SrTiO_3$ and $LaAlO_3$ substrates [28].

**LOW FREQUANCY POLARIZATION RESPONSE.**

Figure 17 presents the temperature dependences of the low-frequency dielectric permittivity $\varepsilon'_c(T,\omega)$ along the *c* axis at several frequencies of the monitoring electric fields. In the studied temperature-frequency region a characteristic narrow peak of the dielectric permittivity is seen at $T_c$ = 27.4 K together with pronounced dispersion in the 70 – 100 K region with increasing in permittivity value at higher temperatures. A drastic change of the dominant character of the dielectric dispersion in the 100-125 K temperature region is seen too. Figure 18 presents the $\varepsilon'_c(T,\omega)$ dependence in the 10 – 85 K region on a larger scale. It clearly shows presence of overlapping high- and low temperature dispersion contributions which form some "background stand" affecting the experimentally measured values of the sharp $\varepsilon'_c(T)$ peak maxima. Obtained picture, presence and the position of the unusually narrow and sharp $\varepsilon'_c(T)$ maximum associated with FE PT are mostly consistent with data [3, 4, 55]. At the same time, authors [4, 55] mainly discussed $\varepsilon'_c$ dispersion at 1 – 100 kHz in the 5 – 60 K region which provides background stand for narrow $\varepsilon'_c(T)$ peak. To our mind, relaxation character of $\varepsilon'_c(T,\omega)$ dispersion, which has been observed in these works, does not connected to behavior of dielectric permittivity in direct neighborhood of $T_c$, i. e. does not connected to relaxation of polarization near second order PT point, which occurs in TMO and DyMnO$_3$ at $T_c = T_{lock}$. Study of the relaxation mechanisms forming the background stand, which at high temperatures is likely mechanism of Maxwell-Wagner (see, e.g. [56]) is beyond the scope of our paper and will not be discussed here.

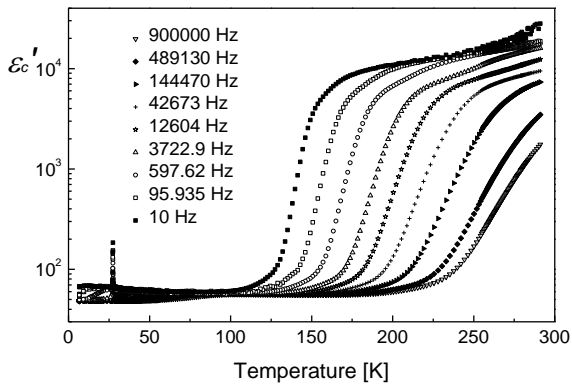
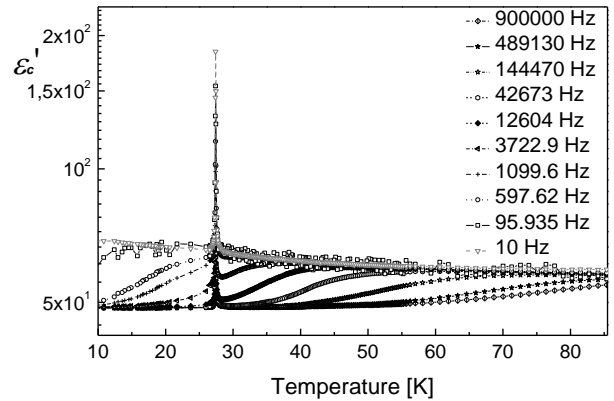

Fig. 17. Dielectric permittivity along *c*-axis ($\varepsilon'_c$) at several frequencies

Fig. 18. Temperature-frequency dependence of $\varepsilon'_c$ in the enlarged scale in the region of 10 – 85 K.

Figure 19 shows behavior of the dielectric permittivity in the region of the narrow $\varepsilon'_c(T)$, peak obtained after separation of the background relaxation contributions. The measurements have been carried out with the temperature resolution of ± 15 mK. It is nicely seen, that in the actual frequency region the magnitude of the $\varepsilon'_c(T)$ maxima increases with the monitoring *ac* field frequency, while any visible change of the temperature of the $\varepsilon'_c(T)$ maximum is not observed experimentally. Such unusual behavior differs significantly from the conventional relaxation picture and the detected features of the low-frequency polarization response require special analysis and explanation.

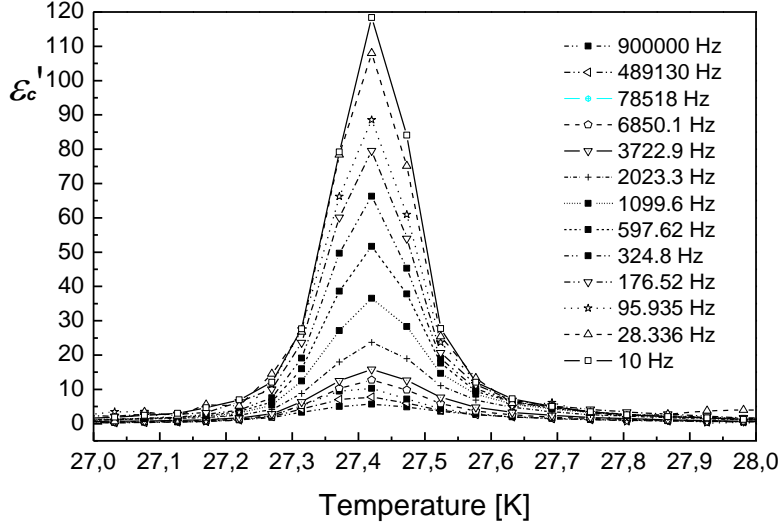

Fig. 19. Dielectric permittivity in the region of narrow temperature peak of $\varepsilon'_c$(T) was taken at different frequencies. For clarity experimental points are connected by lines.

Let us consider the temperature-frequency dependences of the dielectric permittivity in the region of $\varepsilon'_c$(T) temperature maximum in more details. Following to Mostovoj [7] free energy for TMO can be presented as

$$F(\mathbf{M},\mathbf{P}) = \tilde{F}(\mathbf{M}) + \tfrac{1}{2}\chi_0^{-1}P^2 + F_{int}(\mathbf{M},\mathbf{P}); \quad F_{int} = \lambda \mathbf{P} \cdot [(\mathbf{M} \cdot \nabla)\mathbf{M} - \mathbf{M}(\nabla \cdot \mathbf{M})], \qquad (3)$$

where magnetoelectric interaction is described in the form of Lifshitz invariant where constant $\lambda$ is proportional to SOI [7, 9, 10]. Consequently, from the expression (3) for polarization follows $\mathbf{P} = -\lambda\chi_0[(\mathbf{M} \cdot \nabla)\mathbf{M} - \mathbf{M}(\nabla \cdot \mathbf{M})]$. Herewith spin density wave (SDW) $\mathbf{M} = M_b \mathbf{e}_y \cos\mathbf{Q} \cdot \mathbf{r} + M_c \mathbf{e}_z \cos\mathbf{Q} \cdot \mathbf{r} + M_a \mathbf{e}_x$ with wave vector $\mathbf{Q} = (0, Q_b, 0)$ at $M_b M_c \neq 0$ (cycloid structure) is accompanied by homogeneous electric polarization defined by expressions[1]

$$<\mathbf{P}> = \tfrac{1}{V}\int d\mathbf{r}\,\mathbf{P} = (0,0,P_c); \quad and \quad P_c = -\chi_0 \lambda Q_b M_b M_c \ . \qquad (4)$$

From the "magnetic side", contribution of the magnetoelectric interaction into free energy $F_{int}(\mathbf{M},\mathbf{P})$ has pure electronic (spin) nature. At the same time it has dual nature from electric subsystem since electric polarization is a sum of pure electronic contribution $P^{el}$ (at fixed positions of ions) and ionic contribution $P^{ion}$

$$P = P^{el} + P^{ion} \qquad (5)$$

---

[1]*Axes and spatial indices notations correspond to spatial group №62 in $P_{bmn}$ setting. To avoid misunderstanding let us note that in [7] also notations corresponding to standard setting of $P_{nma}$ which differs in orientation of the main axes are used.*

The latter is caused by displacement of ions from high symmetry positions, which, in turn, is caused by the lowering of symmetry of the electron density distribution at magnetic PT. So in the linear approximation $P^{ion}$ can be written as $P_i^{ion} = \frac{e}{v_c}\sum Z_{ij}(s) u_{sj}$, wheree $Z_{ij}(s)$ is effective Born tensor. On the other hand, taking into account that static dielectric susceptibility $\chi_0$ (at $M = 0$) is equal to sum of electronic $\chi_\infty$ and lattice (phonon) $\chi_{ph}$ contributions $\chi_0 = \chi_\infty + \chi_{ph}$, from Eq. (2) and (3) we find that in the linear approximation $P^{el}_c = -\chi_\infty \lambda Q_b M_b M_c$ and $P^{ion}_c = -\chi_{ph} \lambda Q_b M_b M_c$. Thus we obtain that electronic and lattice contributions into spontaneous polarization are connected by simple relationship, which follows already from phenomenology theory: $P^{el}_c$ and $P^{ion}_c$ must have the same sign and its ratio is equal to ratio of the respective susceptibilities, i. e. $P^{el}_c/P^{ion}_c = \chi_\infty/\chi_{ph}$.

From the results of our low-frequency dielectric measurements it follows that for TMO $\chi_\infty/\chi_{ph} \cong 0.1$, from which we obtain that $P^{el}_c \cong 0.1 P^{ion}_c$. Results of *ab initio* calculations [17, 18] yield correct both in sign and in order of magnitude values of $P^{ion}_c$. Slightly worse is situation with electronic contribution $P^{el}_c$, for which both calculations give wrong sign and too small magnitude. The latter is probably caused by the fact that in both works contribution of *f*- electron shell of Tb atoms to polarization factually has not been considered. We also point out that obtained in [18] ratio of polarization magnitudes $P_c(Q_b)$ for two hypothetic modulated structures with $Q_b = 1/3$ and $Q_b = 1/2$: $P_c(1/2)/P_c(1/3) = 1.48$ nicely agrees with result of phenomenological theory $P_c(1/2)/P_c(1/3) = 3/2$ which follows from Eq. (4).

**Low-frequency relaxation near $T_c$**

Behavior of the dielectric susceptibility in the low-frequency region near to Curie point can be described taking into account relaxation dynamic of polarization, which for homogeneous polarization is described by Landau-Khalatnikov equation [34]

$$\frac{\partial}{\partial t} P_c = -\gamma \frac{\partial}{\partial P_c} \Phi(P) = -\gamma \chi_{zz}^{-1}(0) P_c + \gamma E_c, \qquad (6)$$

where $\Phi(P) = F(P) - P \cdot E$, $F(P)$ – is free energy as function of polarization, which appears from Eq. (3) after solving of the magnetic problem, and $\chi_{zz}(0) = \chi_{zz}(\omega = 0, T)$ is static dielectric susceptibility. Passing to Furrier components $P_c(\omega)$ and $E_c(\omega)$, we obtain from Eq. (6) equation for $P_c$: $-i\omega P_c = -\gamma\chi_{zz}^{-1}(0)P_c + \gamma E_c$. From this, taking into account equality $P_c(\omega) = \chi_{zz}(\omega)E_c(\omega)$, we find expression for susceptibility $\chi_{zz}(\omega,T) = (\chi_{zz}^{-1}(0) - i\omega/\gamma)^{-1}$, applicability of which as well as of initial equation (6) in the Landau-Khalatnikov theory is limited by low-frequency region $\omega \ll \gamma$. In [7] it was found that near Curie point $\chi_{zz}(0,T)$ obeys to Curie-Weiss law for $\chi_{zz}(0) \equiv \chi_{zz}(\omega = 0, T)$ with Curie-Weiss constant $C$ is quadratic variable by spin-orbital interaction $C \sim (\lambda\chi_0 Q_b)^2$. In light of the above, it is possible to present real and imaginary part of the dielectric permittivity near to Curie point as:

$$\varepsilon'(\omega,T) = \varepsilon_0 + \frac{4\pi\kappa(T)}{\kappa^2(T)+\left(\frac{\omega}{\gamma}\right)^2}, \quad \varepsilon''(\omega,T) = \frac{\frac{4\pi\omega}{\gamma}}{\kappa^2(T)+\left(\frac{\omega}{\gamma}\right)^2}, \quad \kappa(T) = \begin{cases} 4\pi\frac{T-T_c}{C}, & T \geq T_c \\ 8\pi\frac{T_c-T}{C}, & T < T_c \end{cases} \quad (7)$$

where $\varepsilon_0 = 1 + 4\pi\chi_0$ is dielectric permittivity far from $T_c$. From Eq. (7) it follows that $\varepsilon'(\omega,T)$ has two temperature maxima: in paraelectric phase ($T>T_c$) at $T_{m,>} = T_c + C\omega/4\pi\gamma$ and in FE phase ($T<T_c$) at $T_{m,<} = T_c - C\omega/8\pi\gamma$ with the same magnitude of the both maximums

$$\varepsilon'_{max}(\omega) = \varepsilon'(\omega,T_{m,>}) = \varepsilon'(\omega,T_{m,<}) = \varepsilon_0 + 2\pi\gamma/\omega. \tag{8}$$

Let us consider several characteristic temperature regions important for further discussion of experimental situations:

1) $\Delta T_m = T_{m,>} - T_{m,<}$ - distance between maxima,

2) $\theta_{max}$ – distance at the half of the height between rightmost point $\varepsilon'_>(\omega,T) - \varepsilon_0$ and leftmost point $\varepsilon'_<(\omega,T) - \varepsilon_0$ (width at the half of the height of the double peak $\varepsilon'(\omega,T)$) and

3) $\theta_{min}$ - distance on the half of the height between leftmost point $\varepsilon'_>(\omega,T) - \varepsilon_0$ and rightmost point $\varepsilon'_<(\omega,T) - \varepsilon_0$ (temperature gap on the half of height between two $\varepsilon'(\omega,T)$). Taking into account Eq. (7) we find

$$\Delta T_m = \frac{3C\omega}{8\pi\gamma}, \quad \theta_{max} = (2+\sqrt{3})\Delta T_m, \quad \theta_{min} = (2-\sqrt{3})\Delta T_m. \tag{9}$$

**Comparison with experiment.**

Let us consider $\varepsilon'(\omega,T)$ dependences presented in the Fig. 19 from the point of view of theoretical model developed above. It is seen that It is seen that the main difference is in the number of temperature maxima of $\varepsilon'(\omega,T)$. In the Fig. 19 there is only one permittivity maximum, across the whole $\omega = 10 \div 10^6$ Hz range whereas theory predicts existence of two peaks asymmetric with respect Curie point maxima. According to Eq. (8) and (9) these maxima get closer and grow with the frequency decreasing degenerating in one infinite peak at $T = T_c$ for the static case $\omega = 0$. It follow that frequencies have been used in experiment are too low to observe two peaks of the permittivity, and distance between maxima $\Delta T_m$ in the experimental curve of the Fig. 19 is less than experimental error of determining the temperature, i. e. $\Delta T_m$ satisfies the condition:

$$\Delta T_m = \frac{3C\omega}{8\pi\gamma} \leq \delta_T. \tag{10}$$

From Fig. 3 it is seen that the experimental peaks have anomalously small width: all of them merge with the background at $T < 27$ K and $T > 28$ K. The width at the half-height frequencies (estimation of $\theta_{max}$) is of the order of 0.1– 0.15°, i.e. for the distance between $\varepsilon'(\omega,T)$ maxima this gives (see (9)) $\Delta T_m \approx 0.03°$. Estimation of the same order can be obtained directly from Eq. (9) for $\Delta T_m$ taking into account inequality $\omega \ll \gamma$ and noticing that from the dependence $\varepsilon'(10\text{Hz}, T)$ one can estimate the Curie-Weiss constant as $C \cong 1°$K of the order-of-magnitude. It sets upper bound $\Delta T_m = \frac{3C\omega}{8\pi\gamma} \ll \frac{3C}{8\pi} \approx 0.12°$. These estimations show that for the whole frequency range under study experimental error of the temperature determination $\delta_T \geq 0.01 \div 0.05°$ makes impossible resolution of two maxima of $\varepsilon'(\omega,T)$. An additional difficulty arises from the fact that the minimal distance between wings of two peaks at half height $\theta_{min}$ (the gap between the peaks) is almost 14 times smaller than halfwidth of the double peak, i.e., less than 0.01°K. Taking into account the error of the $\varepsilon'$ magnitude measurement, it hampers drawing of the relief of the experimental curve between two maxima of $\varepsilon'(\omega,T)$. It should be noted that inequality (10) sets the limit on the magnitude of $\gamma$ (at least across the range of frequency diapason $10 - 10^6$ Hz) $\gamma \geq 3\omega C/8\pi\delta_T$. On the other hand, taking into account following from (8) expression $\gamma = \frac{\omega}{2\pi}(\varepsilon'_{max}(\omega) - \varepsilon_0)$, we find that to describe experimental dependences of the Fig. 19 in the framework of this theory the following inequality $\varepsilon'_{max}(\omega) - \varepsilon_0 = 2\pi\gamma/\omega \geq 0.75 \cdot C/\delta_T \approx 0.75/\delta_T \approx 25$ should fulfill. At the same time, the Eq. (8), with regard to inequality $\omega \ll \gamma$ provides direct estimation $\varepsilon'_{max}(\omega) - \varepsilon_0 \gg 2\pi$. These estimations agree well evidencing reasonableness of the theoretical model. At the same time from Fig. 10 it is seen that these estimations hold at $\omega \leq 2$ KHz, and are violated at higher frequencies. Possible cause of discrepancies between theory and experiment is connected, apparently, with the experimental constraints The point is that smallness of the Curie-Weiss constant leads to anomalously small width of the $\varepsilon(0,T)$ curve and the related difficulty of defining with sufficient accuracy as the shape of the $\varepsilon'(\omega,T)$ curves in the low frequency region near the assumed Curie point as the temperature of the $\varepsilon(0,T)$, maximum identified with $T_c$. In particular, this may account for the discrepancy between the behavior of $\varepsilon'(\omega,T)$ at the half-height of the experimental curves are presented in the Fig. 19 and the theoretically predicted (9) linear growth with frequency of the double peak width $\theta_{max}$ at the half-height. Thus, anomalous narrowness of the $\varepsilon'_c(T)$ maximum make impossible observation of the double permittivity maxima in the region of FE phase transition of TMO at frequencies and with accuracy of the temperature determination (~ 15 mK) used in the present work. Note that extension of the frequency in the direction of increase has no sense, because at $\omega \geq$ 1MHz anomalous contribution into $\varepsilon'_c(T,\omega)$ begins to merge with background (see Fig. 19) due to error with determining the background stand of the narrow peak of permittivity and $\varepsilon'_c(T,\omega)$ measurement accuracy.

Concluding the comparison of the theory and experiment, we note that the proposed theoretical model qualitatively correctly describes two most marked features of the observed low-frequency dielectric relaxation associated with the behavior of $\varepsilon'(\omega,T)$ near the Curie point $T_C$: 1) the rapid decrease in the magnitude of the maximum $\varepsilon'_{max}(\omega)$ with increasing frequency $\omega$, and 2) absence of the frequency dependence of the maximum position on the temperature scale.

At the same time, the existing experimental limitations (error in the determination of the temperature, of the background contribution to the $\varepsilon'(\omega,T)$ dispersion, and $\varepsilon'(\omega,T)$) measurements do not allow to observe the true shape of the $\varepsilon'(\omega,T)$ temperature curve with two maxima at high frequencies predicted by Landau-Khalatnikov theory for a critical relaxation of polarization. In this sense, the situation in TMO is turned out to be significantly different from the ferroelectric order-disorder type $Ca_2Sr(C_2H_5CO_2)_6$ with $T_c$ = 283K and $C$=74 K [57] and multiferroic $MnWO_4$ with $T_c$ = 12.6 K [58], in which double permittivity maxima associated with the relaxation of polarization in the region of FE phase transition have been observed at much higher frequencies: of the order ~ 50–100 MHz in $Ca_2Sr(C_2H_5CO_2)_6$ and ~ 200–500 MHz in $MnWO_3$. Besides, extraction of the double permittivity maximum in TMO is significantly hindered due to presence of the "background stands" which are absent as in $MnWO_3$ [58,59] as in $Ca_2Sr(C_2H_5CO_2)_6$ [57].

**Conclusion**

To summarize, we have presented optical ellipsometry and low-frequency dielectric permittivity studies of the spin-spiral-driven ferroelectric-multiferroic TMO single crystals. Detail study of the optical constant and optical absorption spectra in the region of 0.75 – 8 eV have been performed and energies of the main optical transitions are determined. The photoluminescence of TMO have been observed and studied for the first time. A critical analysis of theoretical and experimental data available concerning electronic band structure of TMO was done. Taking into account published data available and energies of the main interband direct and indirect optical transitions determined in our experiment the updated electronic band structure and electronic DOS map of TMO have been suggested. The unknown earlier structured spectra of the optical absorption in the near IR region are found. The lowest indirect optical gap of TMO was found is ~ 0.75 eV. It is shown that thermooptical effect ($n$(T)) for fundamental charge transfer transitions (mostly $p$-$d$) is positive like in the most band dielectrics and semiconductors whereas for low-energy $d$-$d$ transitions, below 40 K, it is negative without visible features in the region of FE phase transition. At the same time, refraction index at $\lambda$ = 495 nm (2.5 eV) reveals a pronounced breakpoint and a rather faint local maximum at ~ 27 K. The temperature minimum of a surface roughness and change of the mechanism of dielectric dispersion were found in the region of 126 K. It can be attributed to presence of the unknown earlier reorganization of the surface structure of TMO, possible, with antiferromagnetic transition due to some change of $Mn^{3+}$ magnetic moment.

A detailed study of the frequency-temperature behavior of the low-frequency dielectric permittivity $\varepsilon'_c(T,\omega)$ in the vicinity of the Curie point $T_c$ = 27.4 K has been performed revealing unusual dispersion of $\varepsilon'_c(T,\omega)$ in the region of a anomalous narrow and sharp peak of the permittivity associated with Curie point. The magnitude of the $\varepsilon'_c(T_c,\omega)$ maximum decreases with frequency, while the temperature of the maximum does not change. Obtained results, analysis and comparability with the published results of other authors indicate that TMO undergo a second order displacive type FE phase transition at $T_c$ = 27.4 K. It was shown that observed peculiarities of the polarization response in the region of sharp $\varepsilon_c'(T)$ maximum and smallness of the $\varepsilon_c'(T)$ magnitude observed in experiment are caused by smallness of the Curie-Weiss constant $C \cong 1$ K and can be described in the frame of Landau-Khalatnikov theory for the critical relaxation polarization..

**Acknowledgements**
V.A.T. is immensely indebted to A.P. Levanyuk and N.E. Christensen for support and valuable discussions. This work was supported in part by the project TA 03001743 of the TACR, 16-22092S of the GACR. The work at Rutgers University was supported by the DOE under Grant No. DOE: DE-FG02-07ER46382.